\documentclass[%
 reprint,
 amsmath,amssymb,
 aps,
]{revtex4-1}

\usepackage{color}
\usepackage{graphicx}
\usepackage{dcolumn}
\usepackage{bm}
\usepackage{nameref,hyperref}

\newcommand{\mean}[1]{\ensuremath{\left< #1 \right>}}

\newcommand{\de}[1]{\textrm{d}#1}

\begin{document}

\preprint{APS/123-QED}

\title{Zipf's and Taylor's laws}




\author{Charlotte James}
\affiliation{Department of Engineering Mathematics, University of Bristol, Woodland Road, BS8 1UB, Bristol, UK}

\author{Sandro Azaele}
\affiliation{School of Mathematics, University of Leeds, Leeds, UK.}

\author{Amos Maritan }
\affiliation{Dipartimento di Fisica ``G. Galilei'', Universit\`a di Padova, INFN, via Marzolo 8, 35131 Padova, Italy.}

\author{Filippo Simini}
\email{f.simini@bristol.ac.uk}
\affiliation{Department of Engineering Mathematics, University of Bristol, Merchant Venturers Building, Woodland Road, BS8 1UB, Bristol, UK}


\date{\today}

\begin{abstract}
Zipf's law states that the frequency of an observation with a given value is inversely proportional to the square of that value; Taylor's law, instead, describes the scaling between fluctuations in the size of a population and its mean. Empirical evidence of the validity of these laws has been found in many and diverse domains. Despite the numerous models proposed to explain the presence of Zipf's law, there is no consensus on how it originates from a microscopic process of individuals dynamics without fine tuning. 
Here we show that Zipf's law and Taylor's law can emerge from a general class of stochastic processes at the individual level, which incorporate one of two features: environmental variability, i.e. fluctuations of parameters, or  correlations, i.e. dependence between individuals. 
Under these assumptions, we show numerically and with theoretical arguments that the conditional variance of the population increments scales as the square of the population and that the corresponding stationary distribution of the processes follows Zipf's law.
\end{abstract}

\pacs{Valid PACS appear here}
\maketitle



\begin{figure}[t]
\centering
\includegraphics[type=pdf,ext=.pdf,read=.pdf,scale = .45]{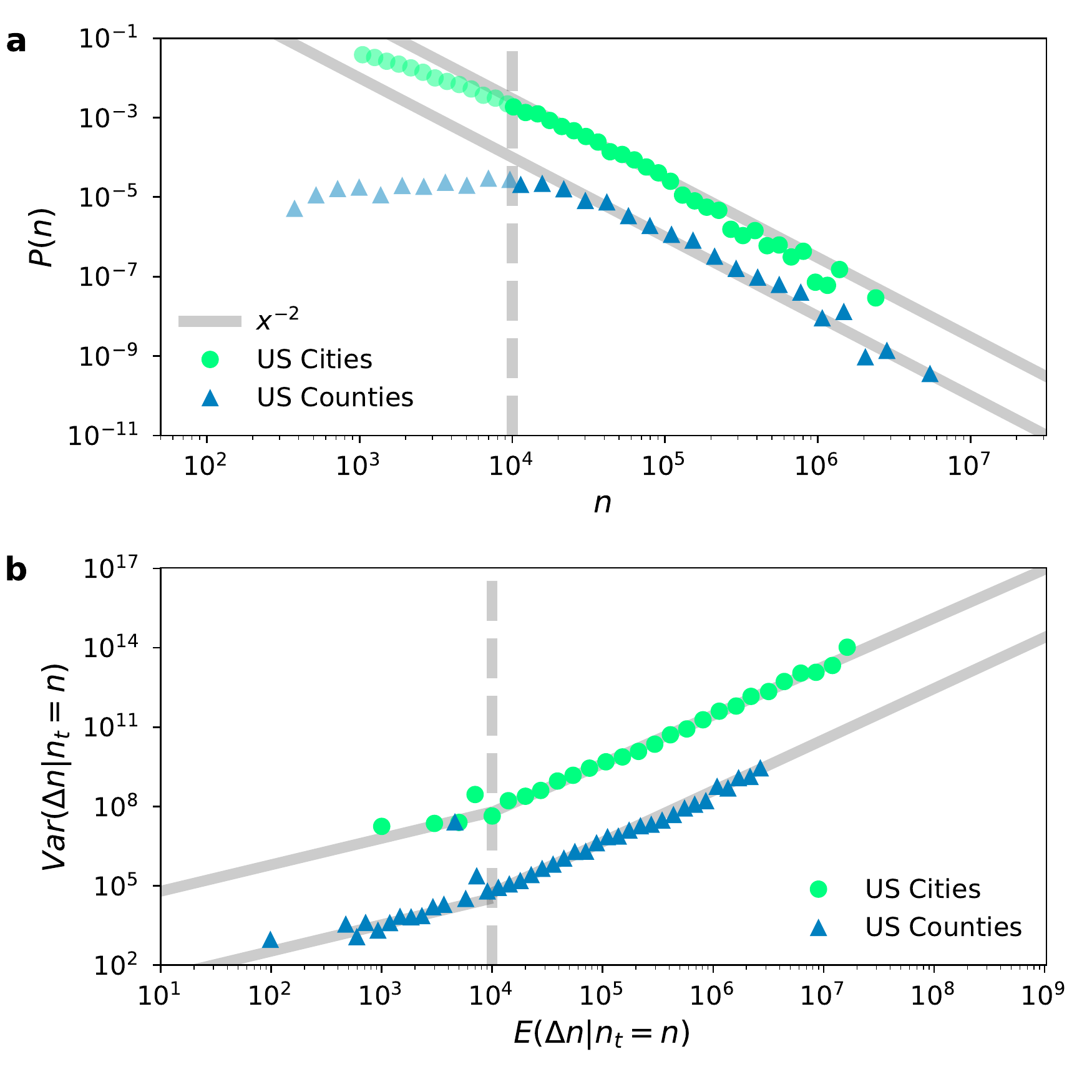}
\caption{
a)  
Zipf's law, data. Probability density function  $P(n)$ (y-axis) vs size $n$ (x-axis) for cities (circles) and counties (triangles) within the United States. 
The solid lines are guides for the eye corresponding to $P(n) \propto n^{-2}$. 
Data on the population of cities is obtained from the {\it Geonames} dataset~\cite{vatant2012geonames}. County level data is obtained from the US Census Bureau~\cite{uscensus}. 
The distribution for US cities has been shifted by a factor of $10$ along the y-axis for clarity.
b) 
Taylor's law, data. The variance of population in year $t+1$ conditioned to the population in year $t$ (y-axis) vs the average population in year $t+1$ conditioned to the population in year $t$ (x-axis) for cities (circles) and counties (triangles) in the United States during the period $1970$ to $2010$. 
The vertical dashed line denotes the cross-over city size $n_{c}$ at which Taylor's exponent defined in Eq. \ref{eq:taylor} transitions from $\alpha = 1/2$ to $\alpha = 1$ (the best fit exponents in the regime $n > n_c$ are $\alpha = 0.92 \pm 0.01$ for cities and $\alpha = 0.97 \pm 0.03$ for counties). As shown in panel a), $n_{c}$ corresponds to the cross-over city size at which the distributions start following Zipf's law. The data for US cities have been shifted by a factor of $10$ along the y-axis for clarity.
\label{fig:data}}
\end{figure}

Formally, a random variable follows Zipf's law if its probability density function is a power law with exponent $-2$. 
Evidence for Zipf's law has been found in many and diverse empirical domains, including systems where observations correspond to groups of individuals and the variable of interest is the group size, such as the number of employees in firms~\cite{axtell2001zipf} or the distribution of family names~\cite{newman2005power}. 
In particular, one of the most documented empirical findings in human geography is Zipf's law for the distribution of city sizes: the probability to find a city with a given 
number of dwellers, $n$, is inversely proportional to the square of that number: 
$P(n) \sim n^{-1-\gamma}$, where $\gamma \simeq 1$. 
The exponent $\gamma \simeq 1$ has been shown to apply to city sizes both globally and historically with surprisingly small deviations. Zipf's law also applies to the distribution of population of larger regions, for example countries in Europe~\cite{giesen2010zipf} as well as cities and counties in the United States (see Fig.\ref{fig:data}a). 

\begin{figure}[t]
\centering
\includegraphics[type=pdf,ext=.pdf,read=.pdf,scale = .45]{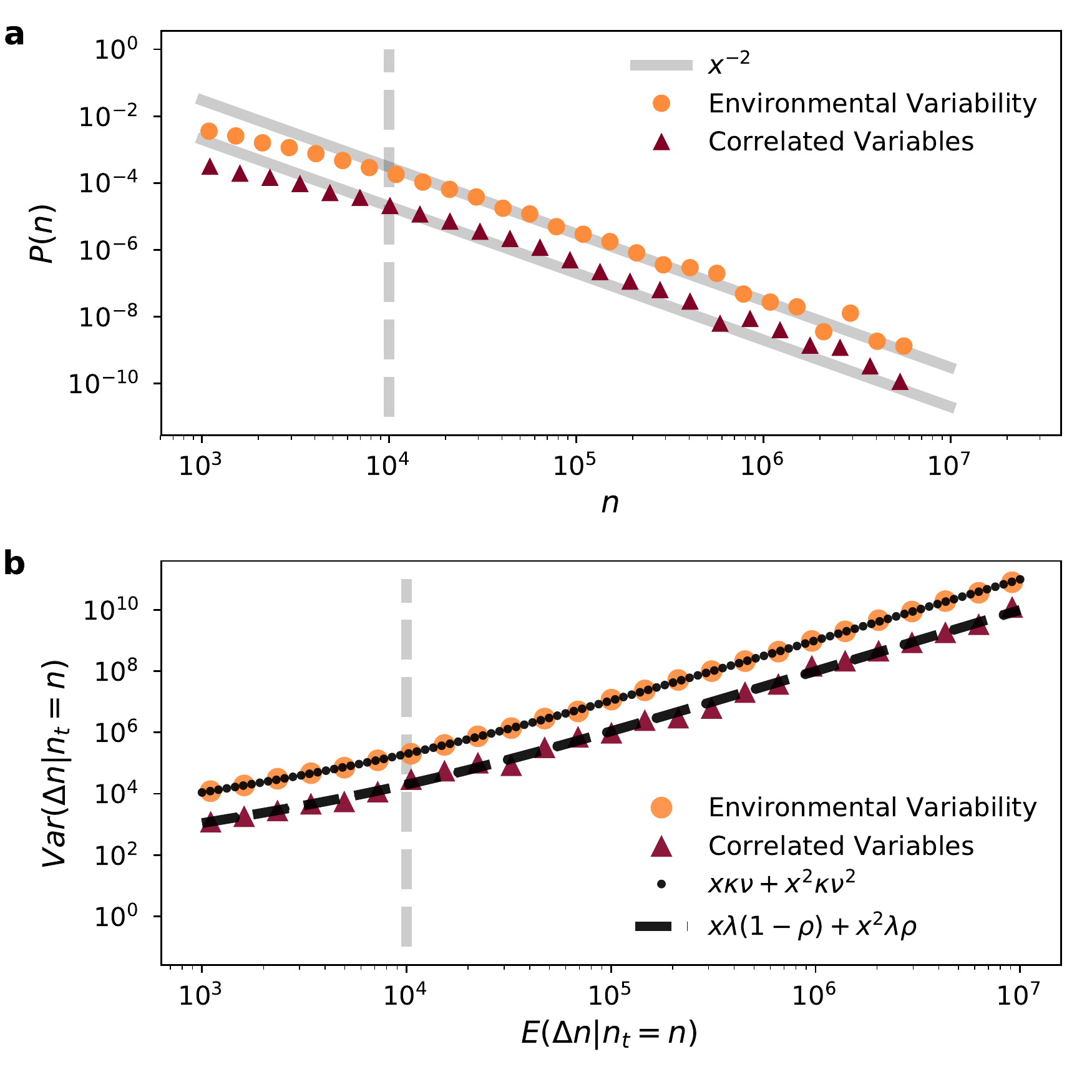}
\caption{ 
a)
Zipf's law, models. 
Stationary distributions of group sizes $P(n)$ for modified branching processes with environmental variability (circles) and correlated individuals (triangles). 
See the main text for details on the numerical simulations and the parameter values. 
The distributions have been shifted by a factor of $10$ along the y-axis for clarity. 
The vertical line denotes the cross-over city size $n_{c}$ at which the distributions start following Zipf's law: $n_c = 10^{4}$ for both models. 
b)
Taylor's law, models. 
The variance of population change in time interval $[t, t+1]$ conditioned to the population at time step $t$ (y-axis) vs the average population change in time interval $[t, t+1]$ conditioned to the population at time step $t$ (x-axis) for branching processes with environmental variability (circles) and correlated variables (triangles). 
Simulations and parameter values are the same as in a). 
We observe a transition of Taylor's exponent from $\alpha = 1/2$ for $n < n_c$ to $\alpha = 1$ for $n > n_c$. The black lines correspond to the analytical results of Eq. \ref{eq:LTV_ev} and Eq. \ref{eq:var_corr}.
Curves have been shifted by a factor of $10$ along the y-axis for clarity.
\label{fig:model}}
\end{figure}

A number of general mechanisms exist to account for the emergence of Zipf's law in various systems~\cite{newman2005power,zanette2005dynamics,
baek2011zipf}. 
A series of recent works~\cite{aitchison2016zipf,schwab2014zipf} has demonstrated that models with latent variables can lead to Zipf's law without fine tuning 
by mixing together narrow distributions with very different means. 
However, these works focus on 
static systems, without an explicit time dependence. 
%
Several models have been proposed to explain how Zipf's law can emerge as the stationary distribution of dynamical processes for the sizes of groups of individuals.  
These models can be divided into two classes based on the scale considered: mesoscopic models at the scale of the groups (e.g., cities) and microscopic models at the scale of the individuals (e.g., dwellers). 
The mesoscopic models are stochastic processes describing the evolution of the group's population as a whole. 
Examples of mesoscopic models are the random multiplicative process~\cite{redner1990random}, also called Gibrat's law \cite{RefWorks:237} or proportionate random growth in the economics literature \cite{RefWorks:279,RefWorks:229}, and those based on the interplay between intermittency and diffusion~\cite{manrubia1998intermittency}. 
%
Mesoscopic models are able to explain the emergence of Zipf's law without the need to fine-tune their parameters to specific values~\cite{gabaix1999zipf}, however they are coarse-grained descriptions of population dynamics and lack an explicit link to the underlying microscopic processes. 
Microscopic models provide a more fundamental description because they are stochastic processes describing the events experienced by an individual, namely births, deaths and migrations, that ultimately determine the change in the size of a population. 
Examples of microscopic models include Yule's and Simon's models based on the rich-get-richer mechanism~ \cite{RefWorks:280,RefWorks:281}, cluster growth and aggregation~\cite{RefWorks:349,makse1998modeling}, preferential migration to large aggregates~\cite{RefWorks:354}, and network growth with redirection~\cite{RefWorks:351}.
However, microscopic models are only able to produce the power law exponent $\gamma = 1$ for specific values of their parameters. 

Here we present 
a class of microscopic stochastic processes that are 
able to reproduce Zipf's law with exponent $\gamma = 1$ without fine tuning. 
These processes are characterised by an anomalous scaling of the 
fluctuations of the population increments, commonly known as Taylor's law in ecology~\cite{eisler2008fluctuation,de2004fluctuations,giometto2015sample}. 
In particular, we describe two general mechanisms to construct microscopic stochastic processes where  
the conditional variance of the population increments scale as the square of the population, and we demonstrate that the stationary distribution of such processes follows Zipf's law. Our processes are applicable to dynamical systems with an explicit time dependence where the stationary distribution of group sizes can be described by Zipf's law. 
In this respect, our derivation of Zipf's law differs from static models without an explicit time dependence and from models where groups can only grow~\cite{aitchison2016zipf,RefWorks:281,zanette2005dynamics, tria2014dynamics,corominas2015understanding}. \\
%
%
We present our microscopic processes as models in which individuals belong to different local areas (groups or patches).
To describe how the population in each group evolves in time, 
we use a modified version of the Galton-Watson process. 
The original Galton-Watson process~\cite{RefWorks:276} is a discrete-time branching stochastic process describing the evolution of a population of $n_t$ individuals at time $t$ according to the equation 
\begin{equation}\label{eq:basic}
n_{t+ 1} = \sum_{i=1}^{n_t} x_i
\end{equation}
where $x_i$ are independent and identically distributed random variables over the integers, with finite mean and variance and with probability mass function $P(x|\lambda)$, where $\lambda$ are parameters. 
If $x_{i} = 0$, individual $i$ dies, if $x_{i} = 1$, they do nothing and if $x_{i} = 2,3,4$... then they have $1,2,3$... children. 
%
Extinction will occur with probability 1 if the average number of offspring per individual is less than or equal to  one, $\mathop{\mathbb{E}}(x) \le 1$. 
To avoid extinction and ensure the process has a stationary state we include the boundary condition $n_t \geq 1$ for all $t$, which accounts for immigration~\footnote{Modelling immigration using a boundary condition is a solution adopted in several models of population dynamics~\cite{azaele2016statistical}. We verified that the specific implementation of the boundary (as a hard reflecting boundary or as a constant influx of individuals) and its value do not affect our results.}.
If $\mathop{\mathbb{E}}(x) > 1$, then the population will experience an exponential growth and there will be no stationary state. 
For $\mathop{\mathbb{E}}(x) \leq 1$, the stationary distribution 
cannot be described by Zipf's law; it is a power law with exponent $-1$ ($\gamma = 0$).

The conditional mean and variance of the population increments, defined as 
$\mathop{\mathbb{E}}(\Delta n | n_t = n)$ 
and 
$Var(\Delta n| n_t = n)$ 
respectively, where $\Delta n = n_{t+1}-n_t$, can be used to measure the scaling of the fluctuations of the population size. 
Given that the random variables $x_i$ are independent and identically distributed, 
the conditional mean and variance of population increments both scale as $n$: 
$\mathop{\mathbb{E}}(\Delta n|n) \sim n$ 
and
$Var(\Delta n|n) \sim n$. 
%
%
This scaling behaviour can be summarised considering the relationship between 
these two quantities, i.e. Taylor's law ~\footnote{Taylor's law originates from ecology, where it was originally defined as the scaling relationship between the variance and mean of populations in different areas~\cite{taylor1961aggregation,giometto2015sample}. Since then, the name ``Taylor's law'' has been often used to denote the general scaling relationship between the variance and the mean of random variables in complex systems~\cite{eisler2008fluctuation,de2004fluctuations}. 
In this case, the random variable of interest is the population increment in one time step, $\Delta n$, and we measure its mean and variance in patches of different sizes, $n$. 
}:
\begin{equation}
Var(\Delta n|n) \propto \mathop{\mathbb{E}}(\Delta n|n) ^{2\alpha}. \label{eq:taylor}
\end{equation}
The exponent $\alpha$ often takes a value of either $1/2$, as in the original Galton-Watson process, or $1$, as 
in random multiplicative processes. 
As random multiplicative processes are known to produce Zipf's law, this suggests Zipf's law can be present with exponent $\gamma = 1$ when Taylor's exponent is $\alpha = 1$. On the other hand we hypothesise that when Taylor's exponent is $\alpha = 1/2$ Zipf's law will not be present. 
This suggests a connection between Taylor's law and Zipf's law, which we characterise with analytical arguments and numerical simulations.

Formalising this intuition, 
we propose two variations of the Galton-Watson process, namely processes with environmental variability and processes with correlated individuals, and show that for large populations they have exponents $\gamma = 1$ and $\alpha = 1$. 
%

%
%
%

\begin{table}[t!!]\centering
\def\arraystretch{1.5}
\begin{tabular}{|c || r | r |r |}
\hline
 \bf Distribution & $\gamma$ & $D$ & $p$ \\
\hline
 US Counties & 0.86 & 0.04 & $<$ 0.01 \\
 \hline
 US Cities & 0.96 & 0.07 & 0.01\\
 \hline
Environmental variability & 0.91 & 0.02 & $<$ 0.01\\
 \hline
Correlated individuals & 1.00 & 0.01 & $<$ 0.01 \\
 \hline
 \end{tabular}
 \caption{Power law fit: exponent $\gamma$ for the distribution of group sizes obtained using a Maximum likelihood estimate, Kolmogorov-Smirnov statistic $D$ and $p$-value for the group size distributions in Figs~\ref{fig:data}a and ~\ref{fig:model}a. All numbers are rounded to two decimal places.}
 \label{tab:results}
 \end{table}

We first present the case of processes with environmental variability. 
%
To this end, we consider 
a modified Galton-Watson process 
where the parameters $\lambda$ of the distribution of the individuals are not constant values but random variables drawn from a distribution $G(\lambda)$ 
at each time step. 
To be specific, 
we assume that the $x_i$ are Poisson random variables with distribution $P(x|\lambda) = \text{Poiss}(\lambda) = e^{-\lambda} \lambda^x / x!$. 



In order to determine the fluctuations of the population increments, we compute the variance of the population change, $\Delta n = n_{t+1} - n_{t}$, conditioned to the population at time $t$ by applying the law of total variance~\cite{weiss2006course}, 
$
Var(\Delta n | n_t = n) = 
\mathop{\mathbb{E}_{\lambda}}[ Var_{\Delta n|\lambda}(\Delta n)] + Var_{\lambda}(\mathop{\mathbb{E}_{\Delta n|\lambda}}[\Delta n])
$.  
Here $\mathop{\mathbb{E}_{\Delta n|\lambda}}$ and $Var_{\Delta n|\lambda}$ denote the mean and variance of $\Delta n$ for a fixed $\lambda$, 
and
$\mathop{\mathbb{E}_{\lambda}}$ and $Var_{\lambda}$ denote the mean and variance with respect to $G(\lambda)$. 
We obtain, 
\begin{equation}\label{eq:LTV_ev}
Var(\Delta n | n_t = n) = 
n \, \mathop{\mathbb{E}_{\lambda}}(\lambda) + 
n^2 \, Var_{\lambda}(\lambda) .
\end{equation}
Note that the fluctuations in the size of a population are proportional to $n$ and 
follow Eq. \ref{eq:taylor} with exponent $\alpha=1/2$ for small populations, 
$n \ll n_c \equiv \mathop{\mathbb{E}_{\lambda}}(\lambda) / Var_{\lambda}(\lambda)$, whereas they scale as $n^2$ for large populations $n \gg n_c$. The crossover population $n_c$ marks the transition between these two scaling regimes. 
Empirical evidence of the presence of this crossover can be found analysing the fluctuations of the populations of cities and counties in the United States, where data shows a transition between Taylor exponents $\alpha = 1/2$ and $\alpha = 1$ around $n_{c} \sim 10^4$ (Fig.~ \ref{fig:data}b). 
Within our framework, this result means that on average the ratio between variance and mean of growth rates is very small, around $n_c^{-1} = Var_{\lambda}(\lambda) / \mathop{\mathbb{E}_{\lambda}}(\lambda) \sim 10^{-4}$.
For example, if the distribution of $\lambda$ is a Gamma distribution, 
$G(\lambda) = \text{Gamma}(\lambda| \kappa, \nu)$, then Eq.~\ref{eq:LTV_ev} becomes 
$Var(\Delta n | n_t = n) = n \, \kappa \nu + n^2 \, \kappa \nu^2$ and 
$n_c = \nu^{-1}$. 
We verify the validity of this prediction 
with numerical simulations of Eq.~\ref{eq:basic} shown in Fig~\ref{fig:model}b, 
where the $x_i$ are independent and identical Poisson random variables with distribution $\text{Poiss}(\lambda)$ and a new parameter $\lambda$ is drawn at each discrete time step from 
a Gamma distribution with fixed parameters $\kappa = 10^4$ and  $\nu = 10^{-4}$. 

Next we demonstrate that the stationary distribution of group sizes follows Zipf's law for large populations, as shown in Fig~\ref{fig:model}a and summarised in Table~\ref{tab:results}.
\begin{figure*}[ht!]
\centering
\includegraphics[type=pdf,ext=.pdf,read=.pdf,scale = .48]{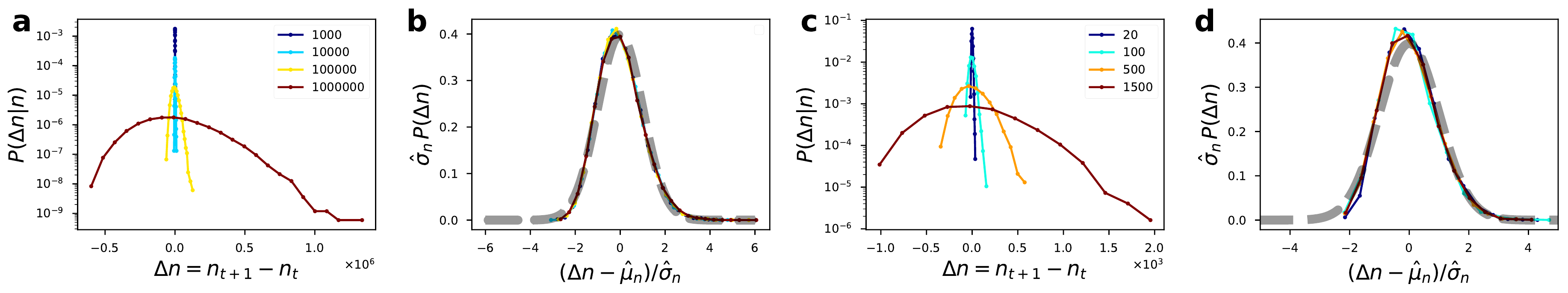}
\caption{  
Numerical simulations support the ansatz of Eq.~\ref{eq:approx_ev} for processes with environmental variability (panels a,b), and with correlated individuals (panels c,d). 
a) Distribution of the fluctuations of the population increments in consecutive time steps, $\Delta n = n_{t+1} - n_t$, for a process with environmental variability where $P(x|\lambda) = \text{Poiss}(\lambda)$ and $G(\lambda) = \text{Gamma}(\lambda| \kappa, \nu)$ with $\kappa = 19.8, \nu = 0.05$. Different curves correspond to different values of populations $n_t = n$. 
b) The curves in panel a) collapse on the same distribution, a standard normal distribution (dashed line), when the population increments are shifted by removing the mean and rescaled by the square root of the variance derived in Eq.~\ref{eq:LTV_ev}. 
c) Distribution of the fluctuations of the population increments in consecutive time steps for a process with correlated individuals, where the $n$ 
Poisson random variables have covariance matrix $Cov(x_i, x_j) = \delta_{i j} \lambda + (1 - \delta_{i j}) \rho \lambda$ with $\lambda = 0.999, \rho = 0.1$. 
 d) The curves in panel c) collapse on the standard normal distribution (dashed line) when the population increments are shifted by removing the mean and rescaled by the square root of the variance derived in Eq.~\ref{eq:var_corr}.  
\label{fig:3}}
\end{figure*}
When the total population is sufficiently large, we expect that Eq.~\ref{eq:basic} can be approximated as 
(see martingale Central Limit Theorem~\cite{Shiryaev_1995}) 
\begin{equation}
\label{eq:approx_ev}
\Delta n \equiv n_{t+1} - n_t \approx \hat{\mu}_n + \hat{\sigma}_n \, \xi(t),
\end{equation}

where 
$\Delta n(t)$ is a continuous random variable, 
$\hat{\mu}_n = \mathop{\mathbb{E}}(\Delta n | n_t = n)$, 
$\hat{\sigma}_n = \sqrt{ Var(\Delta n | n_t = n) }$, 
and $ \xi(t) $ is a zero-mean Gaussian white noise with autocorrelation $ \mean{\xi(t)\xi(t')}=2\delta(t-t') $. 
Figures~\ref{fig:3}a-b demonstrate that numerical simulations support the validity of the ansatz of Eq.~\ref{eq:approx_ev}. 
%
Using the law of total expectation 
and Eq.~\ref{eq:LTV_ev}, we obtain 
$\hat{\mu}_n = n \, (\mathop{\mathbb{E}_\lambda}(\lambda) - 1) \equiv n \hat{\mu}$ and 
$\hat{\sigma}_n = \sqrt{n \mathop{\mathbb{E}_{\lambda}}(\lambda) + n^2 Var_{\lambda}(\lambda)} \equiv \sqrt{n \hat{\sigma}_1^2 + n^2 \hat{\sigma}_2^2}$.
Using the formal substitutions 
$t+1\rightarrow t+\de{t}$,  $\hat{\mu} \rightarrow \mu\de{t}$, $\hat{\sigma_1} \rightarrow \sigma_1 \de{t}$ and $\hat{\sigma_2} \rightarrow \sigma_2 \de{t}$ 
to first order in $ \de{t}$, we obtain the following stochastic differential equation: 
$\dot{n}(t) = \mu n(t) + \sqrt{\sigma_1^2 n(t) + \sigma_2^2 n(t)^2} \xi(t)$, 
with a reflecting boundary at $n = 1$. 
Hence, the proposed birth-death microscopic process can be well approximated by a mesoscopic proportionate random growth dynamics, when populations are large. 
In fact, for large populations, $n \gg n_c = \sigma_1^2 / \sigma_2^2$, the above equation becomes a random multiplicative process with growth rate 
of mean $\mu$ and variance $\sigma_2^2$. 
In this limit, we obtain a stationary distribution with a power-law tail 
$P(n) \sim n^{-2+\frac{\mu}{\sigma_2^2}}$~\cite{van1992stochastic}.  
Notice that we get the exponent of Zipf's law, $\gamma = 1$, for $|\mu| \ll \sigma_2^2$. 
%
The parameter values used in the simulations of Fig.~\ref{fig:model} are chosen to fit the models to city data, capturing both the Zipf exponent of the tail and the large value of $n_c$. 
Because of the specific choice of a Poisson distribution for $P(x|\lambda)$, this requires to use a set of parameter values that are very close to criticality (i.e. $\lambda = 1$). 
However, it is important to emphasize that it is possible to find various distributions $P(x|\lambda)$ and $G(\lambda)$ that produce non-critical systems with stationary distributions with exponent 
$\gamma = -2+\frac{\mu}{\sigma_2^2}$ 
close to Zipf's law for any value of $n_c = \sigma_1^2 / \sigma_2^2$. 
In general, for generic distributions $P(x)$ and $G(\lambda)$, we have:
$$
\begin{cases}
\sigma_1^2    &=   \mathbb{E}_{\lambda} \left(  Var_{\Delta n|\lambda}(\Delta n) \right) \\
\sigma_2^2    &=   Var_{\lambda} \left(  \mathbb{E}_{\Delta n|\lambda}(\Delta n)  \right)  \\
\mu                  &=   \mathbb{E}_{\lambda} \left(  \mathbb{E}_{\Delta n|\lambda}(\Delta n) \right)  - 1 
\end{cases}
$$
so it is possible to find many combinations of functions and parameter values such that 
$|\mu| \ll \sigma_2^2   \ll \sigma_1^2$, corresponding to Zipf's law and large $n_c$. 
For example, it might be possible to satisfy these conditions using a Negative Binomial distribution for $P(x)$, whose mean and variance can be independently set, and choosing an appropriate distribution over its parameters, $G(\lambda)$.
%
%
%
%
%
The second case we present is the class of processes where individuals are correlated. 
To this end, we consider a modified Galton-Watson process with correlated individuals, where the joint probability 
$P(x_1, \dots, x_n| \lambda)$ does not factorize into the product of the individual probabilities $P(x_i| \lambda)$.  To be specific, we assume that the $x$ random variables have a Poisson distribution with fixed parameter $\lambda$ (the same for all individuals) and correlation matrix 
$\rho_{i j} = Cov(x_i, x_j)  / \lambda$, where 
$Cov(x_i, x_j) \equiv \mathop{\mathbb{E}}[(x_i - \lambda)(x_j - \lambda)]$. 
When individuals are correlated the off-diagonal terms of the covariance matrix are non-zero, and in the simplest case that we consider, they are all equal: 
$Cov(x_i, x_j) = \delta_{i j} \lambda + (1 - \delta_{i j}) \rho \lambda$. 
%
%
%
The fluctuations of the population increments for the process with correlated individuals have the same scaling form of the fluctuations for the process with environmental variability. 
The conditional variance of $\Delta n = n_{t+1} - n_{t}$ for a given population $n_t = n$ is: 
\begin{equation}\label{eq:var_corr}
Var(\Delta n | n_t=n) = \sum_{i,j} Cov(x_i, x_j) = 
n \lambda (1 - \rho) + n^2 \rho \lambda 
\end{equation}
Here the cross-over population between the regimes with Taylor exponents $\alpha = 1/2$ and $\alpha = 1$ is 
$n_c = 1/\rho - 1$. 
Within this framework, the cross-over population $n_{c} \sim 10^4$ of United States cities and counties (Fig. \ref{fig:data}b) corresponds to a very small value of correlation between people, $\rho \sim 1/n_c \sim 10^{-4}$. 
The physical meaning of this correlation can be interpreted considering that the inverse of the correlation matrix corresponds to the interaction between individuals~\cite{azaele2010inferring,volkov2009inferring}. Since the off-diagonal elements of the inverse of the covariance matrix vanish in the small $\rho$ limit we get that only a small amount of interaction is needed for large group sizes to follow Taylor's law with exponent $\alpha = 1$, yet in the absence of correlations this will not be present. 

Using the ansatz of Eq.~\ref{eq:approx_ev} and following the same steps taken for the case with environmental variability, it is possible to show that the process with correlated individuals has a stationary distribution of group sizes which for large populations, $n > n_c$, is a power-law, 
$P(n) \sim n^{-2 + \frac{\lambda-1}{\rho \lambda}}$, that
follows Zipf's law when $|\lambda-1|/\lambda \ll \rho$. 
Numerical simulations support the validity of the approximation of Eq.~\ref{eq:approx_ev} on the scaling of the population increments (Figs.~\ref{fig:3}c-d and Fig.~\ref{fig:model}b) and the presence of Zipf's law in the distribution tail (Fig.~\ref{fig:model}a, Table~\ref{tab:results}).

In the numerical simulations, we used two methods to generate correlated Poisson variables. 
In the first method, used in the simulations of Fig.~\ref{fig:model}, we consider the modified Galton-Watson process 
$n_{t + 1} = \sum_{i=1}^{n_t} (x_i + z)$
where the $x_i$ are independent and identical random variables with Poisson distribution $\text{Poiss}(\lambda(1 - \rho))$ and we introduce the random variable $z$, independent of the $x_i$, with Poisson distribution $\text{Poiss}(\rho \lambda)$. 
One can show that for this process the conditional variance of $\Delta n$ is identical to Eq.~\ref{eq:var_corr}. 
In Fig.~\ref{fig:model} we use $\lambda = 1$ and $\rho = 10^{-4}$.  
%
In the second method, used in the simulations of Fig.~\ref{fig:3}c-d, we use a Gaussian copula model to link the Poisson marginals~\cite{barbiero2015simulation}. 
We verified with numerical simulations that for appropriate values of the multivariate Gaussian's parameters, this method generates Poisson variables with the desired values of $\lambda$ and covariance matrix $Cov(x_i, x_j) = C_{i j}$; 
the conditional variance of $\Delta n$ is again identical to Eq.~\ref{eq:var_corr}.
Both methods give results compatible with the theoretical predictions.

To summarise our results, we have shown that 
environmental variability and correlations between individuals are able to produce Zipf's law and Taylor's law~\footnote{Indeed, numerical simulations show that the same results can be obtained when both mechanisms are present.}. 
The exponent of Taylor's law 
shifts from $\alpha = 1/2$ for small populations to $\alpha = 1$ for large populations. 
Also, when population is large 
Zipf's law emerges naturally without fine tuning whenever the growth rate's mean is much larger than the variance. 
The proposed framework can also naturally account for exponent values close but not identical to Zipf's law. 
This reveals a general connection between Zipf's law and Taylor's law in microscopic stochastic processes of population dynamics under realistic assumptions.

\begin{acknowledgments}
FS is supported by EPSRC First Grant EP/P012906/1.
\end{acknowledgments}

\bibliography{individual}

\end{document}